\documentclass[journal]{IEEEtran}

\ifCLASSINFOpdf
\else
   \usepackage[dvips]{graphicx}
\fi
\usepackage{url}

\usepackage{graphicx}
\usepackage{bm}
\usepackage{multirow}
\usepackage{booktabs}
\usepackage{makecell}
\usepackage{bbding}
\usepackage{balance}
\usepackage{float}
\usepackage{amsmath}
\usepackage{amsfonts}
\usepackage{bm}

\begin{document}

\title{Interactive Dual-Conformer with Scene-Inspired Mask for Soft Sound Event Detection}

\author{Han Yin, \IEEEmembership{Student Member, IEEE}, Jianfeng Chen, \IEEEmembership{Senior Member, IEEE}
\thanks{This paragraph of the first footnote will contain the date on which you submitted your paper for review. It will also contain support information, including sponsor and financial support acknowledgment. For example, ``This work was supported in part by the U.S. Department of Commerce under Grant BS123456.'' }
\thanks{The next few paragraphs should contain the authors' current affiliations, including current address and e-mail. For example, F. A. Author is with the National Institute of Standards and Technology, Boulder, CO 80305 USA (e-mail: author@boulder.nist.gov).}
\thanks{S. B. Author, Jr., was with Rice University, Houston, TX 77005 USA. He is now with the Department of Physics, Colorado State University, Fort Collins, CO 80523 USA (e-mail: author@lamar.colostate.edu).}}

\markboth{Journal of \LaTeX\ Class Files, Vol. 14, No. 8, August 2015}
{Shell \MakeLowercase{\textit{et al.}}: Bare Demo of IEEEtran.cls for IEEE Journals}
\maketitle

\begin{abstract}
Traditional binary hard labels for sound event detection (SED) lack details about the complexity and variability of sound event distributions.
Recently, a novel annotation workflow is proposed to generate fine-grained non-binary soft labels, resulting in a new real-life dataset named MAESTRO Real for SED. In this paper, we first propose an interactive dual-conformer (IDC) module, in which a cross-interaction mechanism is applied to effectively exploit the information from soft labels. In addition, a novel scene-inspired mask (SIM) based on soft labels is incorporated for more precise SED predictions.
The SIM is initially generated through a statistical approach, referred as SIM-V1. 
However, the fixed artificial mask may mismatch the SED model, resulting in limited effectiveness.
Therefore, we further propose SIM-V2, which employs a word embedding model for adaptive SIM estimation.
Experimental results show that the proposed IDC module can effectively utilize the information from soft labels, and the integration of SIM-V1 can further improve the accuracy. 
In addition, the impact of different word embedding dimensions on SIM-V2 is explored, and the results show that the appropriate dimension can enable SIM-V2 achieve superior performance than SIM-V1.
In DCASE 2023 Challenge Task4B, the proposed  system achieved the top ranking performance on the evaluation dataset of MAESTRO Real.

\end{abstract}


\IEEEpeerreviewmaketitle

\section{Introduction}
\label{sec:intro}

\IEEEPARstart{S}{ound} event detection (SED)  involves automatically identifying specific sound events and providing their temporal locations from acoustic signals. 
SED-related systems are applied in various applications such as medical surveillance [1] and smart home automation [2].

Recently, many neural network-based methods for SED are proposed, such as convolutional neural network (CNN) [3], recurrent neural network (RNN) [4] and convolutional recurrent neural network (CRNN) [5].
In addition, the convolution-augmented transformer (Conformer), has also been widely employed for more accurate SED [6,7]. 
Conventionally, strongly-labeled datasets are used to train above-mentioned SED models, in which textual labels, onsets and offsets are provided for the sound event instances.

In traditional strong labels for SED, each sound event is assigned a binary activity level, indicating occurrence (0) or non-occurrence (1). 
Such \textit{hard labels} are advantageous for creating unambiguous training categories, but lack some details about certain intricacies of event distribution [10].
In complex and variable real acoustic environments, the probability of a sound event occurring can not be simply represented by binary values [11].
Recently, a novel annotation workflow has been proposed to produce more reliable and objective strong labels for SED [12].
The strong labels obtained are called \textit{soft labels}, which contain non-binary values representing the probability of event occurrence.
Compared to traditional binary hard labels, soft labels can provide additional fine-grained information.
Based on such novel annotation workflow, a new real-life dataset named MAESTRO Real is generated for SED with soft labels[17].
During evaluation, these soft labels are converted to hard labels using a fixed threshold.

Some works have been studied to utilize the information within soft labels, trying to improve the performance of SED.
For example, CRNN-based models are used to extract the beneficial information contained within soft labels [13].
In addition, the pre-trained model, audio spectrogram transformer [14], is applied as a feature extractor to generate frame-wise embeddings for training based on soft labels [15].
Recently, the spectro-temporal receptive field is incorporated in convolutional layers to build a human auditory soft SED system [16].
However, information from soft labels still can not be effectively utilized by the above methods, resulting in limited performance on MAESTRO Real.

\begin{figure*}
\centering
\centerline{\includegraphics[width=0.9\textwidth]{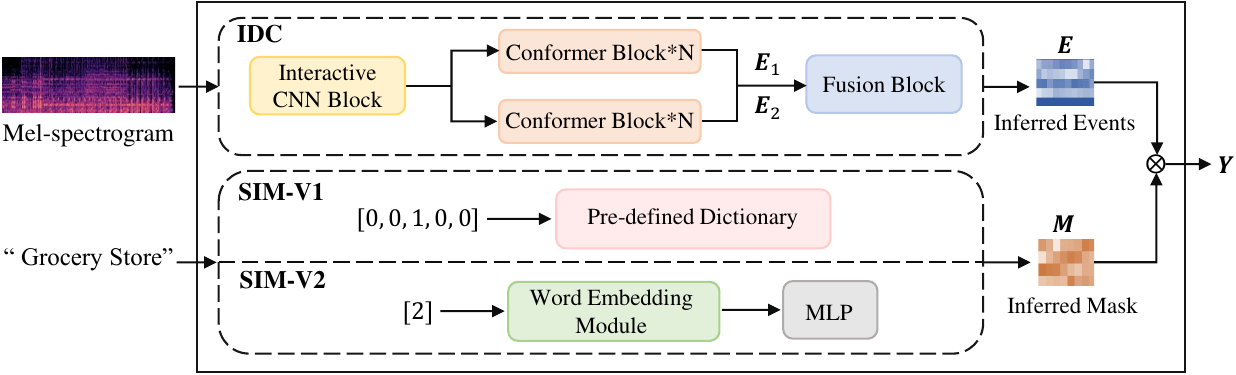}}
\caption{The overview of proposed interactive dual-conformer (IDC) with scene-inspired mask (SIM) for soft sound event detection, where SIM-V1 and SIM-V2 are two different approaches for mask estimation.}
\label{fig:overview}
\end{figure*}

In this letter, we first present an \textbf{i}nteractive \textbf{d}ual-\textbf{c}onformer (IDC) module, in which a cross-interaction mechanism is applied to exploit the rich information in soft labels. 
Furthermore, considering that acoustic scenes and sound events are explicitly related, a novel \textbf{s}cene-\textbf{i}nspired \textbf{m}ask (SIM) based on soft labels is incorporated for more precise SED predictions.
This integration leads to a notable enhancement on the performance of SED.
Specifically, the proposed SIM contains the probabilities of various sound events occurring under different acoustic scenes, 
which is initially generated through a statistical approach, referred as SIM-V1.  
However, such fixed artificial mask may mismatch the SED model, resulting in limited detection performance. 
Therefore, we further propose SIM-V2, which employs a word embedding module for adaptive SIM estimation, achieving superior performance than SIM-V1. 
Proposed systems have achieved the state-of-the-art (SOTA) performance on the evaluation dataset of MAESTRO Real.

\section{Methods}
\label{sec:method}
In this section, we first present the architecture of the proposed IDC module and the two different SIM estimation approaches.
In Sec.\ref{sec:loss}, we explain how to effectively exploit the information from soft labels based on the IDC module.

\subsection{Interactive Dual-Conformer Module}
\label{sec:IDC}
As shown in Fig.\ref{fig:overview}, the IDC module consists of three components: the interactive CNN block, the Conformer block and the fusion block. 
Details are described as follows.
\subsubsection{The Interactive CNN Block}
\begin{figure}
\centering
\centerline{\includegraphics[width=0.4\textwidth]{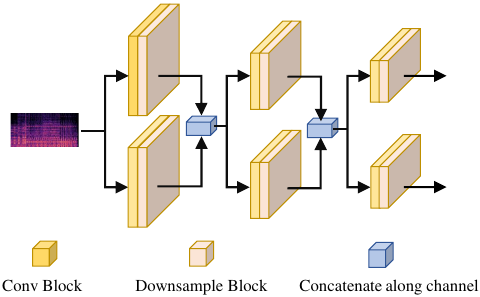}}
\caption{The architecture of the proposed interactive CNN block.}
\label{fig:idc}
\end{figure}

As shown in Fig.\ref{fig:idc}, this block is a dual-path structure. 
Convolutional layers of different branches can extract diverse characteristics from the mel-spectrogram.
In addition, feature maps from different branches are concatenated along the channel axis to achieve information interaction.
By introducing this cross-interaction mechanism, the network can better capture the correlation between different features, thereby improving the performance and generalization ability of the model.
Specifically, the Conv Block consists of a convolution layer, a batch normalization layer and a ReLU activation function. And the Downsample Block is composed of a max pooling layer and a dropout layer. We do not perform any pooling operations in the time dimension to obtain frame-level sound event detection estimation.
Detailed configurations are presented in Table \ref{tab:config}. 

\subsubsection{The Conformer Block} 
The Conformer block [1] is originally proposed to address speech recognition problems, 
which employs a combination of the multi-head self-attention mechanism and the convolutional structure to better model long-distance dependencies.
This structure can adapt to inputs of different lengths or frequencies and achieve superior performance on other speech or audio-related tasks [1-5].
Therefore, we propose to utilize Conformer blocks downstream for dynamic audio information extraction.
We do not change the structure of the Conformer block, but only add a fully connected (FC) layer and a sigmoid activation function at the end to generate sound event estimations.

\subsubsection{The Fusion Block} 
As shown in Fig.\ref{fig:overview}, the goal of this block is to perform adaptive weighting of frame-level sound event estimations $\bm{E}_{1} \in \mathbb{R}^{T \times K_e}$ and $\bm{E}_{2} \in \mathbb{R}^{T \times K_e}$ generated by different branches, where $T$ and $K_e$ represent the number of frames and event categories respectively. 
The process can be formulated as:
\begin{equation}
    \bm{E} =\bm{E}_1(i,j)\cdot\bm{\mu}(j)+\bm{E}_2(i,j)\cdot[1-\bm{\mu}(j)]
\end{equation}
where $i=1,2,...,T-1$, $j=1,2,...,K_e-1$, and $\bm{\mu} \in \mathbb{R}^{K_e}$ is the learnable vector.
This mechanism performs adaptive weighting based on the sound event categories and is adaptable to audio of varying durations, 
making it more computationally convenient during inference.

\subsection{Scene-Inspired Mask V1 and V2}
\label{sec:SIM}
In the real-world environment, sound events and acoustic scenes are explicitly related.
Some sound events have a high probability of occurring in specific scenes and vice versa.
For example, ``metro approaching'' is almost impossible to happen in a grocery store.
This scene-event pattern can be represented as a dictionary $\bm{D} \in \mathbb{R}^{K_s \times K_e}$, where $K_s$ represents the number of scene categories.
The values in the dictionary range from 0 to 1, indicating the probability of various sound events occurring in each scene.

Based on this scene-event dictionary, a frame-level mask $\bm{M} \in \mathbb{R}^{T \times K_e}$ can be generated that contains the probabilities of different events for each frame.
At the end of the model, the mask and the SED prediction are element-wise multiplied to further improve accuracy, which can be formulated as:
\begin{equation}
    \bm{Y} = \bm{E}\odot\bm{M}
\end{equation}
Since the mask $\bm{M}$ is produced based on the acoustic scene information, we refer it as the scene-inspired mask, denoted as SIM.
As shown in Fig.\ref{fig:overview}, two different approaches are presented for SIM estimation, details are described as follows.

\subsubsection{SIM-V1}
This is a completely statistical method, resulting in a fixed SIM. 
Firstly, we count the probability of various events occurring in different scenes based on soft labels.
And the scene-event dictionary can be defined by:
\begin{equation}
    \bm{D}(m,n) = \sigma_{m,n}
\end{equation}
where $m=1,2,...,K_s$ and $n=1,2,...,K_e$, $\sigma_{m,n} \in [0,1]$ represents the statistical probability of the $n$th type of event occurring in the $m$th type of scene.

Then, the frame-level SIM can be generated based on the predefined dictionary, as formulated in:
\begin{equation}
    \bm{M} = \bm{1}\cdot\bm{\lambda}\cdot\bm{D}
\end{equation}
where $\bm{1}\in \mathbb{R}^{T \times 1}$ is all-one vector, and $\bm{\lambda} \in \mathbb{R}^{1 \times K_s}$ is the one-hot vector converted from the scene of the input audio. 

\subsubsection{SIM-V2}
Different from SIM-V1, this approach aims to leverage the word embedding module and the multi-layer perceptron (MLP) to perform adaptive estimation.
Assume that the input audio belongs to the $k$th type of scene, and the dimension of the word embedding model is $K_h$. 
We first use the word embedding module to produce the frame-level embedding $\bm{H} \in \mathbb{R}^{T \times K_h}$.
Then, the MLP is utilized to bulid non-linear projections from $\bm{H}$ to the final SIM.
The whole process can be formulated as:
\begin{equation}
\begin{aligned}
    &\bm{H} = {\rm Embed}(\bm{1}\cdot k) \\
    &\bm{M} = f_2[{\rm ReLU}(f_1(\bm{H}))]
\end{aligned}
\end{equation}
where $\bm{1}\in \mathbb{R}^{T \times 1}$ is all-one vector, $f_1(\cdot)$ and $f_2(\cdot)$ are both FC layers, 
in which the number of hidden units is ${K_h/2}$.

In SIM-V2, the scene-event dictionary is optimized through gradient descent and is saved among the parameters of the word embedding module,
which can be generated by:
\begin{equation}
    \bm{D} = f_2[{\rm ReLU}(f_1(\bm{W}))]
\end{equation}
where $\bm{W}\in \mathbb{R}^{K_s \times K_h}$ is the weight of the word embedding module.
Therefore, the size of $K_h$ has an important impact on the performance of SIM-V2.

\subsection{Loss Functions}
\label{sec:loss}
As described in Sec.\ref{sec:intro}, compared with traditional binary hard labels, soft labels contain more fine-grained information. 
However, during evaluation, soft labels still need to be converted to binary hard labels through a fixed threshold.
Denote the soft label as $\bm{Y}_s \in \mathbb{R}^{T \times K_e}$, the hard label $\bm{Y}_h \in \mathbb{R}^{T \times K_e}$ can be generated by:
\begin{equation}
    \bm{Y}_h(p,q)=
    \begin{cases}
    1&\bm{Y}_s(p,q)\geq\mu\\0&\bm{Y}_s(p,q)<\mu
    \end{cases}
\end{equation}
where $p=1,2,...,T$, $q=1,2,...,K_e$ and $\mu$ is the fixed threshod.
The transformed hard label can be regarded as the discretized representation of the coarse-grained information contained in the soft label.

The coarse-grained information mainly focuses on overall semantic information, while the fine-grained information provides more detailed local features about sound events. 
By leveraging both types of information, semantic understanding of the audio can be improved, leading to more accurate detection performance.
Therefore, based on the proposed IDC module, the loss function can be formulated as:
\begin{equation}
    \begin{aligned}
        {\rm Loss} &= {\rm BCE}(\bm{E}_1, \bm{Y}_h) + {\rm MSE}(\bm{E}_2, \bm{Y}_s)\\
                     &+ {\rm BCE}(\bm{E}, \bm{Y}_h) + {\rm MSE}(\bm{E}, \bm{Y}_s)
    \end{aligned}
\end{equation}
where ${\rm BCE}(\cdot)$ and ${\rm MSE}(\cdot)$ are the binary cross entropy and the mean square error functions respectively.

Furthermore, we present two additional loss functions as baselines, formulated as:
\begin{equation}
    {\rm Loss}_A = {\rm BCE}(\bm{E}, \bm{Y}_h)
\end{equation}
\begin{equation}
    {\rm Loss}_B = {\rm MSE}(\bm{E}, \bm{Y}_s)
\end{equation}
In ${\rm Loss}_A$, only the coarse-grained hard labels are utilized, while ${\rm Loss}_B$ only uses the fine-grained original soft labels.

\section{Experiments}
\label{sec:exp}
In this section, we present details about the dataset and the training process.
Experimental results are then discussed to demonstrate the effectiveness of proposed methods.
\subsection{Dataset and Evaluation Metrics}
All experiments are  conducted on MAESTRO Real, which is the only publicly available soft-labeled dataset for SED.
The dataset consists of real-life recordings with a length of approximately 3 minutes each, recorded in 5 different acoustic scenes: cafe restaurant, grocery store, city center, residential area and metro station.
There are 11 categories of sound events as follows: 
birds singing,
car,
people talking,
footsteps,
children voices,
wind blowing,
brakes squeaking,
large vehicle,
cutlery and dishes,
metro approaching and 
metro leaving. 

This dataset is used in the DCASE 2023 Challenge Task4B, divided into the development set and the evaluation set.
The audio and corresponding labels in the development set are both publicly available, with a total duration of 190 minutes.
While the evaluation set only provides 97 minutes of audio, labels are utilized by the DCASE committee.
And the evaluation is based on the following metrics, all calculated in 1s-segments:
\begin{itemize}
\item[*] $F1_{mi}$: micro-average F1 score with a threshold of 0.5
\item[*] $ER_{mi}$: micro-average error rate with a threshold of 0.5
\item[*] $F1_{ma}$: macro-average F1 score with a threshold of 0.5
\item[*] $F1_{mo}$: macro-average F1 score with the optimal threshold for each class
\end{itemize}

\subsection{Training strategy and Model configuration}
128-dimensional log-mel spectrogram is extracted with a window size of 500 ms and a hop size of 250 ms.
During training, we feed audio with a fixed duration of 25 s to the model.
Models are trained with a 5-fold crossevaluation setup same as the official DCASE baseline.
The number of the Conformer block is set to $N=2$.
Configurations of the proposed interactive CNN Block are presented in Table \ref{tab:config}.

Adam optimizer is employed to update weights with a batch size of 32.
Dropout rate is set to 0.1 and
learning rate is initialized to 0.001, which is automatically halved when there is no performance improvement for 10 epochs continuously. 
Training stops when learning rate drops 5 consecutive times.

\begin{table}[htbp]
\centering
\caption{Configurations of the proposed interactive CNN Block}
\renewcommand\arraystretch{1.5}{
\setlength{\tabcolsep}{3.3mm}{
\begin{tabular}{ccc}
\toprule[2pt]
Block                                 & Filter size@Filters        & Output       \\
\midrule[1pt]
input                                 & -                          &(1, 100, 128)  \\
Conv Block1                           & (3$\times$3)@32            &(32, 100, 128) \\
Downsample Block1                     & (1$\times$5)               &(32, 100, 25)  \\
Conv Block2                           & (3$\times$3)@64            &(64, 100, 25)                      \\
Downsample Block2                     & (1$\times$2)               &(64, 100, 12)                      \\
Conv Block3                           & (3$\times$3)@128           &(128, 100, 12)                     \\
Downsample Block3                     & (1$\times$2)               &(128, 100, 6)                     \\
\bottomrule[2pt]
\end{tabular}
}
}
\label{tab:config}
\end{table}

\subsection{Experimental Results}
\subsubsection{Ablation Study}
Firstly, different loss functions are considered to guide the training of proposed networks to obtain an optimal strategy.
Table 2 shows the performance of the IDC system trained with three loss functions described in Sec.\ref{sec:loss} on the development datdaset.
$Loss(\cdot)$ outperforms other loss functions in all metrics, 
indicating that better detection performance can be achieved by using both coarse-grained and fine-grained information from soft labels simultaneously.

\begin{table}[htbp]
\centering
\caption{Results of the IDC system trained with three different loss functions on the development dataset}
\renewcommand\arraystretch{1.5}{
\setlength{\tabcolsep}{3.3mm}{
\begin{tabular}{ccccc}
\toprule[2pt]
Loss Function                        & $F1_{mi}$   & $ER_{mi}$   & $F1_{ma}$   & $F1_{mo}$    \\
\midrule[1pt]
$Loss_A(\cdot)$                       & & & & \\
$Loss_B(\cdot)$                       & & & &\\ 
$Loss(\cdot)$                         & & & &\\
\bottomrule[2pt]
\end{tabular}
}
}
\label{tab:loss}
\end{table}

Then, to demonstrate the effectiveness of SIM, we compare the performance of different systems trained with the same loss function on the development set.
As shown in Table \ref{tab:SIM}, compared with IDC, the $ER_{mi}$ of IDC-SIM-V1 decreased by $11.5\%$ and $F1_{mo}$ increased by $3.5\%$, 
which shows that the integration of SIM-V1 can improve the accuracy of SED estimations.
When we replace SIM-V1 with SIM-V2, which is learned from the 64-dimensional word embedding module, the overall performance of SED drops slightly.
As the word embedding dimension increases, the performance of SIM-V2 gradually improves, and the embedding dimension of 256 enables SIM-V2 $1.86\%$ higher than SIM-V1 in $F1_{mo}$.
This is because compared to the fixed SIM-V1, SIM-V2 is an adaptive estimation method that can progressively adapt to the output of the IDC module during the training process.

\begin{table}[htbp]
\centering
\caption{Results of different systems trained with the same loss function on the development dataset}
\renewcommand\arraystretch{1.5}{
\setlength{\tabcolsep}{2mm}{
\begin{tabular}{cccccc}
\toprule[2pt]
System             &\begin{tabular}[c]{@{}c@{}}Embedding\\ Dimension\end{tabular}               & $F1_{mi}$   & $ER_{mi}$   & $F1_{ma}$   & $F1_{mo}$    \\
\midrule[1pt]
IDC              &-         & & & &\\
\hline 
IDC-SIM-V1       &-         & & & &\\
\hline
\multirow{3}{*}{IDC-SIM-V2}    &$\bm{K}_h=64$              & & & &\\
\cline{2-6}
                  &$\bm{K}_h=128$ & & & &\\ 
\cline{2-6}
                  &$\bm{K}_h=256$ & & & &\\ 

\bottomrule[2pt]
\end{tabular}
}
}
\label{tab:SIM}
\end{table}

\subsubsection{Comparison With the SOTA Systems}

\begin{table}[htbp]
    \centering
    \caption{Results of the proposed systems and other SOTA systems on the evaluation set (embedding dimension is 256 in IDC-SIM-V2)}
    \renewcommand\arraystretch{1.3}{
    \setlength{\tabcolsep}{3.3mm}{
    \begin{tabular}{ccccc}
        \toprule[2pt]
        System   & $F1_{mi}$   & $ER_{mi}$   & $F1_{ma}$   & $F1_{mo}$ \\
        \midrule[1pt]
        CRNN                & 74.13\%       & 0.484     & 35.28\%       & 43.44\%        \\
        H. Zhang et al.     & 77.86\% & 0.367 & 35.71\% & 43.60\%   \\
        T. D. Nhan et al.             &- & - &- & 47.17\% \\   
        D. Min et al. 	    & 78.05\%	& 0.361 & 29.19\% &	48.95\%   \\
        X. Xu et al. 	    & 80.80\%	& 0.329 & 35.58\% &	51.13\%   \\
        M. Chen et al. 	    & 80.89\%	& \textbf{0.320} & 31.74\% &	52.03\%   \\
        \midrule[0.6pt]
        IDC	                & & & &\\
        IDC-SIM-V1	        & & & &\\
        IDC-SIM-V2	        & & & &\\
        \bottomrule[2pt]
    \end{tabular}
    }}
    \label{tab:SOTAs}
\end{table}

\section{Conclusion}

A conclusion section is not required. Although a conclusion may review the main points of the paper, do not replicate the abstract as the conclusion. A conclusion might elaborate on the importance of the work or suggest applications and extensions. 



\begin{thebibliography}{34}
\bibitem{}G. O. Young, ``Synthetic structure of industrial plastics,'' in {\em Plastics}, 2nd ed., vol. 3, J. Peters, Ed. New York, NY, USA: McGraw-Hill, 1964, pp. 15--64.

\bibitem{}W.-K. Chen, {\it Linear Networks and Systems}. Belmont, CA, USA: Wadsworth, 1993, pp. 123--135.

\end{thebibliography}
\end{document}